\documentclass[fleqn,usenatbib]{mnras}
\usepackage{graphicx}
\usepackage{url}
\usepackage{longtable}
\usepackage{float}
\usepackage{subfigure}
\usepackage{newtxtext,newtxmath}
\usepackage[T1]{fontenc}
\usepackage{amsmath}
\usepackage{soul}
\usepackage{lscape}

\title{Density Exponent Analysis: Gravity-driven steepening of the density profiles of star-forming regions }
\author[Li \& Zhou]{
  Guang-Xing Li,$^{1}$\thanks{gxli@ynu.edu.cn, ligx.ngc7293@gmail.com} 
Ji-Xuan Zhou$^{1}$ \\
$^{1}$ South-Western Institute for Astronomy Research, Yunnan University, Chenggong District, Kunming 650091, P.\,R. China}
\begin{document}

\maketitle
% \affil{South-Western Institute for Astronomy Research, Yunnan University, Kunming 650500, People's Republic of China}
% \begin{affiliations}
%  \item South-Western Institute for Astronomy Research, Yunnan University, Kunming 650500, People's Republic of China
% %  \item separate with \verb|\item| commands.
% \end{affiliations}
  % Shaped by processing including turbulence and gravity, the density structure of molecular clouds are highly complex and inhomogeneous, which poses challenges to analysis   

\begin{abstract}
 
  The evolution of molecular interstellar clouds is a complex, multi-scale process.   The power-law density exponent describes the steepness of density profiles, and it has been used to characterize the density structures of the clouds yet its usage is usually limited to spherically symmetric systems.
  Importing the Level-Set Method, we develop a new formalism that generates robust maps of a generalized density exponent $k_{\rho}$ at every location for complex density distributions. 
  By applying it to high fidelity, high dynamical range map of the Perseus molecular cloud constructed using data from the Herschel and Planck satellites, we find that the density exponent exhibits a surprisingly wide range of variation ($-3.5 \lesssim
  k_{\rho} \lesssim -0.5$). Regions at later stages of gravitational collapse are associated with steeper density profiles.
  Inside a region, gas located in the vicinities of dense structures has very steep density profiles with $k_{\rho} \approx -3$, which forms because of depletion. 
  This density exponent analysis reveals diverse density structures, forming a coherent picture that gravitational collapse leads to a continued steepening of the density profile. We expect our method to be effective in studying other power-law-like density structures, including granular materials and
  the Large-Scale Structure of the
  Universe. 
\end{abstract}

  
\section{Introduction}
Many astrophysical processes, including the gravitational collapse of molecular
 clouds, are complex and multi-scaled. Residing in the Galactic disk, the clouds
 are the nurseries of stars. They are open systems that interact with the
 environment constantly. Their collapses involve an interplay between turbulence
 \citep{2004RvMP...76..125M}, gravity, magnetic field \citep{2014prpl.conf..101L},
 ionization radiation, and Galactic shear, resulting in highly complex density
 distributions.

 \begin{figure}
  \includegraphics[width = 0.45   \textwidth]{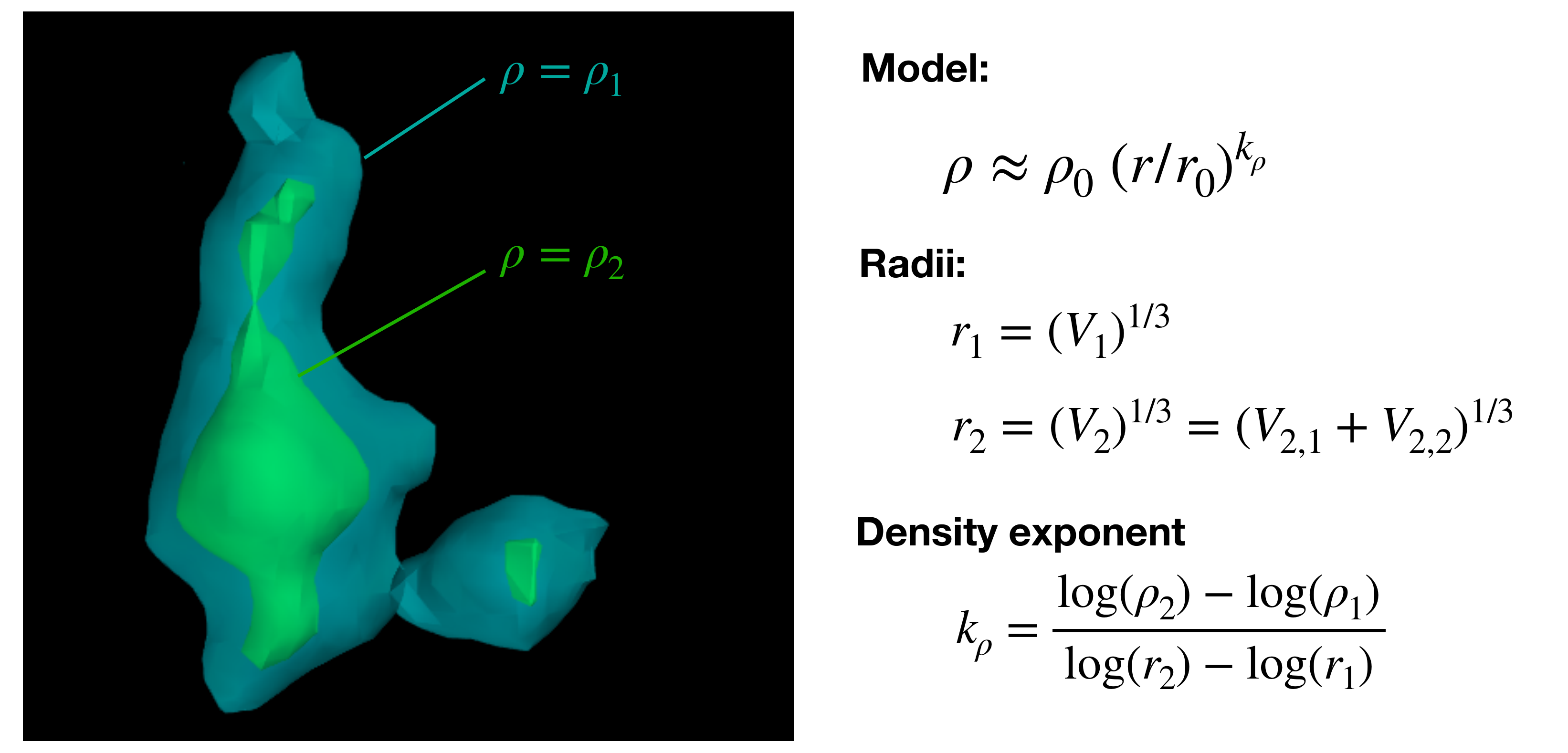}
  \caption{\label{fig:illus}  {\bf Evaluation the Level-Set Density Exponent.}
  We first divide a region using a set of isosurfaces. A typical region
  $R_1$ would be surrounded by an isosurface at $\rho = \rho_1$. Inside this
  region, there exists one or a few subregions  ($R_{2, i}$) surrounded by isosurfaces
  at  $\rho = \rho_2$. The equations on the right hand side describe how the
  density exponent $k_{\rho}$ at voxels included in $R_1$ yet not included in
  $R_{2, i}$) is computed. The spacing between the adjacent
  isosurfaces are exaggerated for a clearer view.  }
\end{figure}

 A clear picture of how the collapse occurs is yet to be achieved. Modern, high
 spatial dynamical range observations provide maps that contain an unprecedented
 amount of information \citep{2016A&A...587A.106Z}. Measures like density
 (Probability Distribution Function) PDF \citep{2009A&A...508L..35K},
 correlation function \citep{2003ApJ...588..881P} and fractal dimension
 \citep{1993A&A...270..451H}  and spectral correlation function \citep{1999ApJ...524..887R}
have been proposed to quantify the structure of the
 star-forming regions and these are summarized in \citet{2021PASP..133j2001B}
 . Despite their successes, these measures are degenerate,
 where original data from regions are often represented using curves where the spatial information is lost. Although the density PDF has enjoyed good success
 in quantifying the cloud structure  and connecting simulations with observations
 \citep{2009A&A...508L..35K,2011ApJ...727L..20K,2012ApJ...750...13C,2016MNRAS.461.3027L,2021MNRAS.507.4335K},
 the loss of information during the compression means that the complexity of the
 underlying structures is often overlooked. Besides, to derive these measures,
 one needs to specify the boundary of a region in advance, which can be a
 a challenging task and efforts have been made to address the boundary issue
 \citep{2015A&A...578A..97L,2015IAUGA..2253648A} The criteria for collapse
 remains an open question. Addressing the transition from clouds to clumps which
 hosts clustered star formation, the virial parameter
 \citep{1992ApJ...395..140B} is often used to study whether a region should
 collapse or not, and gravitationally unbound clouds like the Perseus  can be divided into regions which gravitationally bound by themselves
 \citep{2015A&A...578A..97L}. Along this direction, \citep{Li2017} provides a
 criteria describing the transition to gravitationally-bound clumps. Analytical
 models have been established to describe the evolution of density structure and
 the star formation efficiencies in a turbulent medium
 \citep{2005ApJ...630..250K,2007ApJ...661..972P,2015MNRAS.450.4035F,2017ApJ...834L...1B,2019MNRAS.488.1501G}.

 Power-law density structures such as  $\rho \propto r^{k_{\rho}}$ ($k_\rho <
 0$) are common in systems that have reached intermediate asymptotic states,
 such that the behaviors are independent of the details of the initial and/or
 boundary conditions \citep{1996sssi.book.....B,Goldenfeld}. One such example is
 the gravitational collapse of a molecular cloud, where stationary collapse leads
 to $\rho\sim r^{-2}$ \citep{1969MNRAS.144..425P,
 2009ARep...53.1127P,2011ApJ...727L..20K,Girichidis2014,2015ApJ...814...48N,2015ApJ...804...44M,2018MNRAS.474.5588D,2018MNRAS.477.4951L,2019MNRAS.485.3224D},
 and accretion flow around dense objects  have $\rho\sim r^{-1.5}$
 \citep{1941MNRAS.101..227H,1952MNRAS.112..195B}. For these systems, measuring
 the power-law density exponent $k_\rho$ would enable us to distinguish
 different structures, and the value of $k_{\rho}$ can be directly compared
 against models to achieve understanding.  Various attempts have been made to
 measure the density exponent.  The most obvious approach is to fit spherical
 models to observational data. However, as the majority of regions we study are
 non-spherical, this approach is limited in practice.  Another way is to derive
 the power-law density exponent using the density probability distribution
 function (PDF) \citep{Girichidis2014,2016MNRAS.461.3027L,2020ApJ...903...56P}.
 Although the procedure is straightforward,  this statistical approach only
 allows for the derivation of an ``effective'' density exponent, which contains
 no information on how gas organizes spatially.

% In previous studies

% the  derive the density exponent by fitting power-law models to data, or drive an ``effective'' density exponent using the surface density PDF. In the former case, the model usually assumes spherical density disbituion, which is overly simplified, and in the latter case, the ``effective" density e version of the 

% In this paper, we propose a new Level-Set Method-based formalism to dissect ana analyze the structure of these regions. 
To fully exploit the diagnostics power of the density exponent $k_{\rho}$, importing the Level-Set Method (LSM),
 we propose a new formalism to measure its value for non-spherical yet centrally condensed regions. The Level-Sets are contourlines, and the Level-Set Method is a conceptual framework where analyses of
surfaces and shapes can be performed with the help of Level-Sets. 
By applying the LSM to state-of-the-art high dynamical range observations of star-forming regions,  we obtain spatially-resolved maps of the density exponent, and reveal, for the first time, the complexity and regularity of molecular cloud structures. 
% Our new results point to a unified model where gravitational collapse and accretion lead to a continuous steepening of the density exponent. 

% From an observational point of view, measuring $k_{\rho}$ in different regions offers a new way to distinguish between different regions. In spite of the usefulness,  evaluating $k_{\rho}$ observationally can be a challenging task, as the model has assumed a symmetric density structure that is overly simplified compared to the very complex structure we observed. 

% We propose that the concept of density exponent $k_{\rho}$ can be generalized to smooth yet complex density distributions to quantify the steepness of the density profile measured in the log-log space. Our method is inspired by the Level-set method (LSM), where our regions divided with isosurfaces such that a generalized version of the density exponent $k_{\rho}$  can be evaluated using adjacent isosurfaces. 
\begin{figure*}
  \centering
  \includegraphics[width = 0.9   \textwidth]{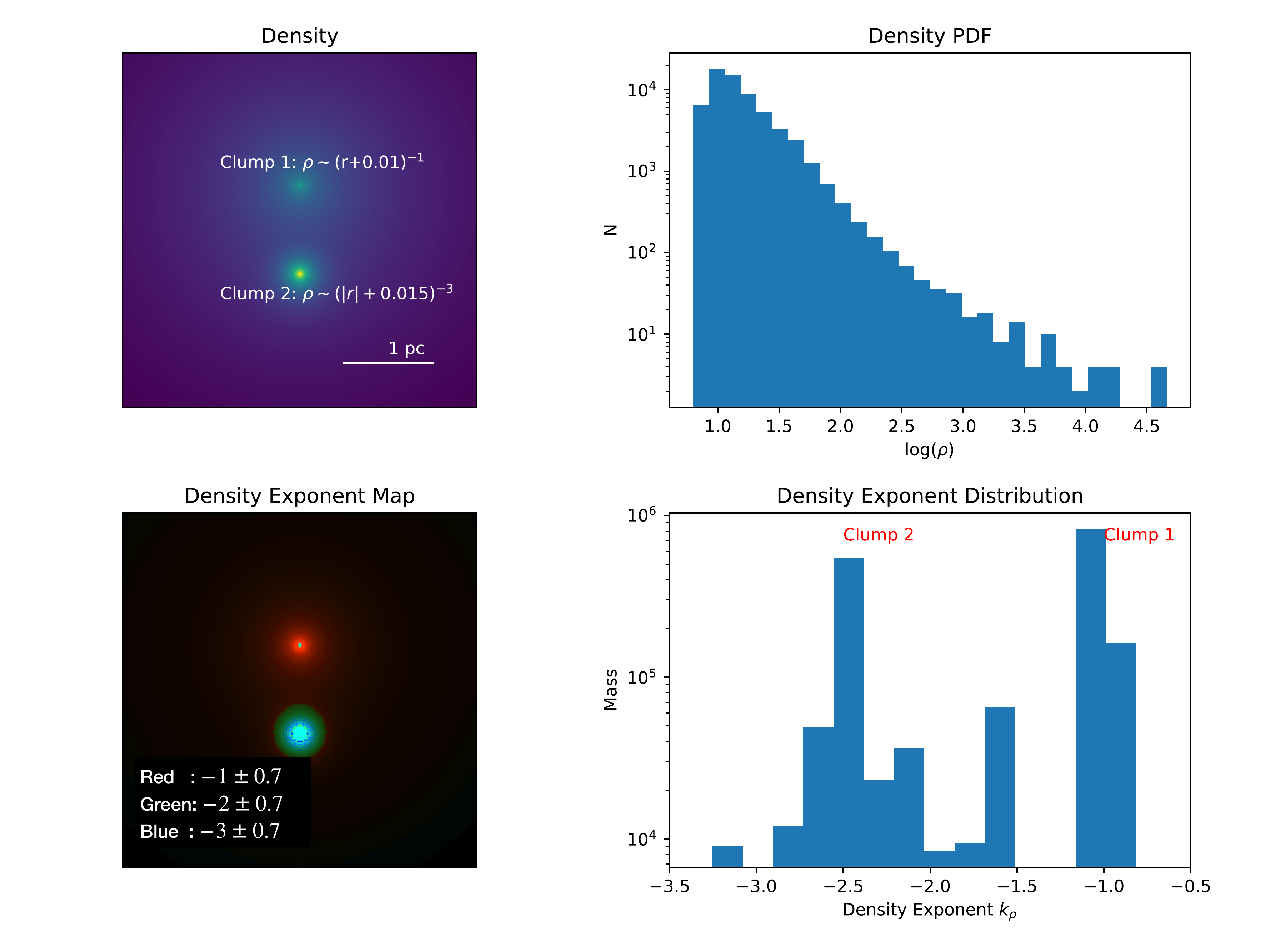}
    \caption{\label{fig:illus}  {\bf Methods for analyzing cloud structure.} We consider an example made of two clumps (clump 1: $\rho \sim (r + 0.01 )^{-1}$, clump 2: $\rho \sim (r + 0.015)^{-3}$). Because of the truncation at small radii, the density exponent of the clump 2  is larger than -3.  
    {\bf Upper Left:} Density distribution. {\bf Upper Right:} Density PDF. {\bf Lower Left:} Density Exponent Map. Color represents the density exponent, and  brightness represents surface density.  The colormap is the same as the one in Fig. \ref{fig:overview}.    {\bf Lower Right:} mass-weighed density exponent distribution. The normalization of the $y$-axis is arbitrary. \label{fig:0}}
\end{figure*}

\section{Method}
% {\bf Level-Set Density Exponent:}
For a spherical system, the density structure can be described as $\rho(r)$ where $r$ is the radius. To measure the steepness of the density profile in a given location, we  adopt a local model  where $\rho  \sim r ^{k_{\rho}}$, and in the vicinity of $r$, the value of $k_{\rho}$ can be derived as 
$$
k_{\rho}(r) = \frac{{\rm log}(\rho(r + \delta_r))- {\rm log}(\rho(r))}{ {\rm log}(r + \delta_r) - {\rm log}(r) }\;,
$$
where $\delta_r << r$.
  
The goal is to measure the density exponent for clouds of arbitrary geometries. Assuming a 3D density structure $\rho(x, y, z)$,
   we divide the region using a
  set of densely-spaced iso-density contours, after which each subregion should be surrounded by
  a contour at $\rho = \rho_1$, one or a few contours at $\rho = \rho_2$, and it
  should contain values ranging from $\rho_{1}$ to $\rho_{2}$. Assuming
  that the  region $R_1$ has a volume of $V_1$, inside this region, there
  can be a few ($n$) subregions $R_{2, i}$ surrounded by isosurfaces with $\rho
  = \rho_{2}$ ($\rho_2 > \rho_1$), and these subregions have volumes of $V_{2,
  i}=V_{2,1}, \ldots V_{2,n}$. The size of the region can be approximated as $r_1
  \propto (V_1)^{1/3}$, and the effective size of all subregions altogether can be
  approximated as $r_2 \propto ({\sum_i V_{2, i}})^{1/3}$. The \textit{Level-Set Density Exponent} is
\begin{equation}
  k_{\rho} =  \frac{{\rm log}(\rho_2) - {\rm log}(\rho_1)}{{\rm log}(r_2) - {\rm log}(r_1)} \;.
\end{equation}
The procedure 
is illustrated in Fig. \ref{fig:illus}, and the resulting map is called the \emph{Density Exponent Map}.

% In the spherical case, the Level-Set Density Exponent is reduced to the ``traditional'' density exponent for spherical regions. For non-spherical regions, the level-set method provides a generalized way to define the size and the density exponent.
% After obtaining the Level-Set Density Exponent, we write its values back to the corresponding regions to produce a map of $k_{\rho}$. This procedure is named as ``Density Exponent Analysis''. 
The advantage of the Density Exponent Analysis lies in its robustness and resolving power:
 the method is directly applicable to maps that contain heterogeneous structures and can be used to distinguish these structures.
As an example (Fig. \ref{fig:0}), we construct a model which contains two spherical clumps of different density profiles. We derive its density PDF and produce a density exponent map. The density PDF  contains limited information since the spatial information is lost completely.  In contrast, the Density Exponent Map indicates that the map should be separated into regions characterized by different density exponents and the map contain values of the density exponent at every location. The additional spatial information retrieved by the method makes it a powerful tool to analyze complex, spatially inhomogeneous datasets, such as the density structures of nearby molecular clouds. We should note that our resolved map of the density exponent is not completely identical to the density exponent derived in analytical models  \citep{1969MNRAS.144..425P,
2009ARep...53.1127P,2011ApJ...727L..20K,Girichidis2014,2015ApJ...814...48N,2015ApJ...804...44M,2018MNRAS.474.5588D,2018MNRAS.477.4951L,2019MNRAS.485.3224D}, as in those cases, the density fluctuations caused by e.g. turbulence are not accounted for.

A final remark: one can be temped to evaluate the density exponent in 2D, where 
\begin{equation}
  k_{\Sigma} =  \frac{{\rm log}(\Sigma_2) - {\rm log}(\Sigma_1)}{{\rm log}(r_2) - {\rm log}(r_1)} \;,
\end{equation}
where $\Sigma$ represents the surface density, and derive $k_\rho$ using $k_{\rho} = k_{\Sigma} - 1$. However, this approach is limited, and the above-mentioned equation holds only if the structures we study are completely self-similar \footnote{\url{https://en.wikipedia.org/wiki/Abel_transform}.}.

\section{Results \&  Discussions}
% The best way to realize the power of the method is to apply it to high-fidelity, high-spatial dynamical range observations of nearby star-forming regions. 
Located at a distance of $\sim$ 290 pc
\citep{2018ApJ...869...83Z}, the Perseus star-forming region is nearby and well-resolved.
We use the surface density map derived using data from the
Herschel and the Planck telescope \citep{2016A&A...587A.106Z}. The original map is the opacity of dust at 870 $\mu$m. We converted the map into a map of the surface density using a calibrated conversion factor provided in their Appendix. 
 The
cloud has a size of $\sim$ 30 pc, and the map has a resolution of 36 arcsec ($\sim
0.05\;\rm pc$). The enormous  ($\sim 10^3$) spatial dynamical range allows detailed studies. 

Derivation of the level-set density exponent $k_{\rho}$ requires 3D density
distributions. As observations are done mostly in 2D, we develop a method (see Method \ref{sec:appen:3d})
to construct 3D density distributions using 2D maps. This is achieved by first decomposing a 2D surface density map into component maps that contain structures of different sizes and by assigning thicknesses to these component maps
and combining them. For cloud-like density structures, the
reconstruction allows us to measure the mass-weighted mean density exponent to an
accuracy of $\lesssim 0.1$. 
In our calculations, we focus on gas with
$\rho_{\rm H_2} > 1000 \;\rm cm^{-3}$.
 This corresponds to  40\% of the gas contained in the Perseus region and the region is surrounded by a diffuse envelope that contains gas that does not contribute directly to the star formation. We also excluded unresolved regions -- patches surround by contours whose sizes are smaller than 3 pixels (0.08 pc), from our analyses. 
 To derive the density exponent map, the data is divided using 100 contour levels equally spaced in log($\rho$).

%  We also excluded gas contained in density peaks whose sizes
% are smaller than 3 pixels, as the still-limited resolution  of the map  prevents us from
%  evaluating the density exponent on these scales. f

\begin{figure*}
  \includegraphics[width = 1.0 \textwidth]{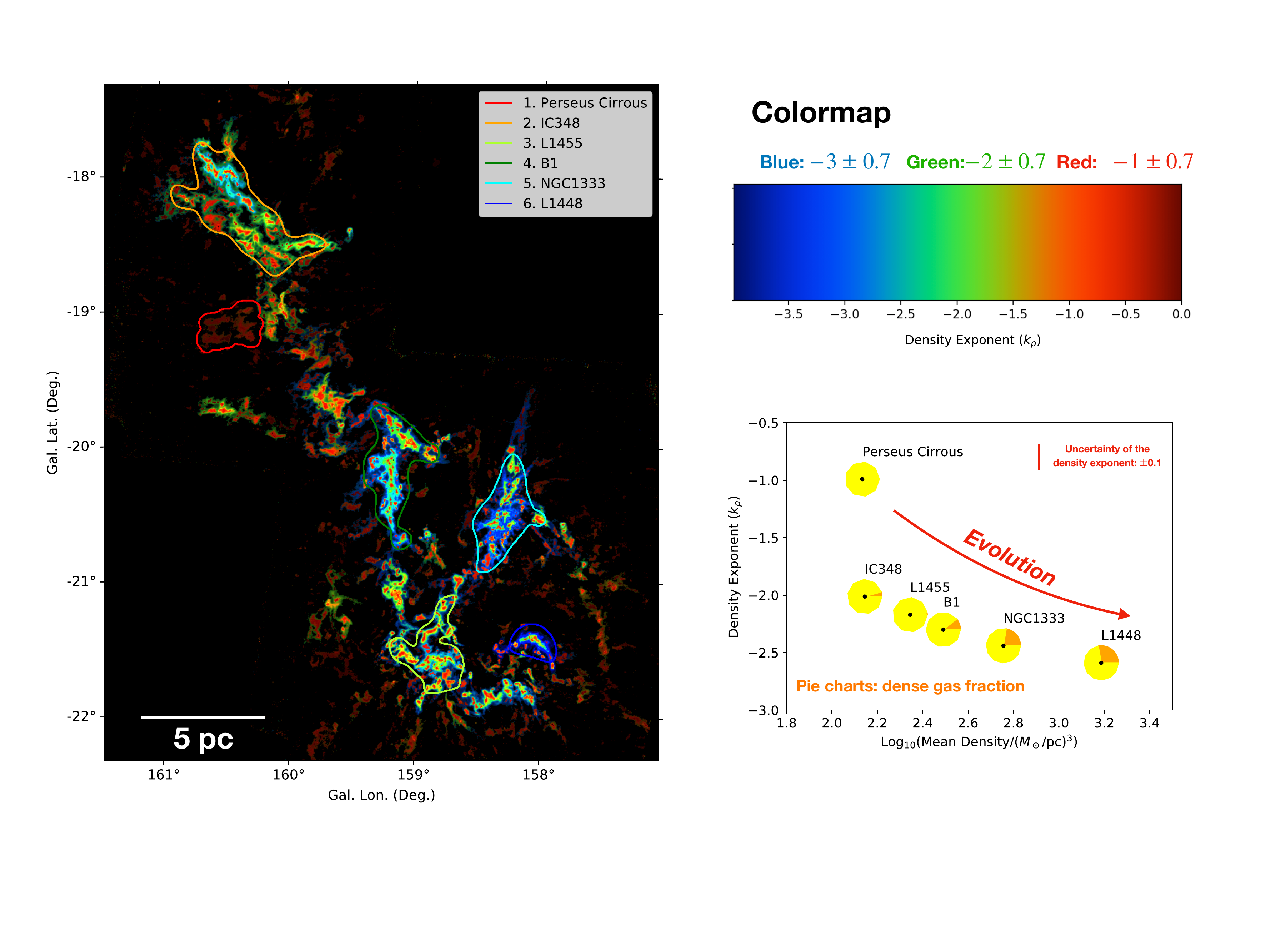}
  \caption{ {\bf Results from the Perseus molecular cloud.} 
  {\bf Left:}  
Density Exponent Map of the Perseus region. We plot the results at the midplane of the reconstructed 3D density distribution.  The brightness represents the density, and the colors represent the density exponent $k_{\rho}$.  The color channels have  response functions of Gaussian shapes (see the colormap on the upper right). 
% Red color represent regions where $k_{\rho} = -1 \pm 0.7 $, green color represent regions where $k_{\rho} = -2 \pm 0.7 $, and blue color represent regions where $k_{\rho} = -3 \pm 0.7 $. 
% The ranges of $k_{\rho}$ different colors represent are chosen to overlap slightly to ensure a smooth look. 
  {\bf Right:} Density exponent plotted against the mean density.  Pie charts show the dense gas mass fractions of the corresponding regions, where the yellow areas stand for the diffuse gas and the orange areas stand for the dense gas.  A relation between dense gas fraction and density exponent can be found online. For a single region, the mass-weighted mean density exponents has an estimated uncertainty of 0.1, caused by our density reconstruction (See our online material). 
   \label{fig:overview} }
\end{figure*}
In Fig. \ref{fig:overview}  we plot the density exponent distribution at the cloud center plane, which contains all the line-of-sight density maximums. $k_{\rho}$ ranges from -3.5 to -0.5, forming a highly inhomogeneous pattern that contains variations on different scales. 
We note that when the structures we study can be characterized by strict power laws, the slope of density PDF is $3/k_\rho $, and the slope of the surface density PDF is  $2/(k_\rho + 1)$. In our case, the range of $k_{\rho}$ translates to density PDF slopes that range from -1 to -6, and a surface density PDF slopes as shallow as -1. We note that these are very rough translations, as the derivations hold only if the underlying structures are strictly self-similar.

\subsection{Inter-regional variations: Density-driven collapse}
The Perseus clouds can be separated into a few pc-sized subregions. Each region has a corresponding gravitational potential dip \citep{2015A&A...578A..97L}, and these regions can collapse to form star clusters or associations. We first divide the cloud into these regions and evaluate parameters including the mass-weighed mean density exponent and the mean densities. 
  We also characterize these regions by deriving a quantify called the dense gas fraction (see Methods \ref{sec:appendix:dense}). Since all the dense gas would collapse to form individual or multiple stars, the dense gas mass fraction is a direct indicator of the star formation activity.

Correlations between these quantities are summarized in the right panel of Fig. \ref{fig:overview}.
In general, regions of steeper density profiles have higher dense gas mass fractions. Thus,  the formation of dense gas, which occurs on tiny scales, is deeply linked to a global steepening of the density profile. Although the link between dense gas mass fractions and the steepness of the density profile inferred from the shape of the surface density PDF is somehow established \citep{2015A&A...577L...6S,2016MNRAS.461.3027L}, our maps provide a more detailed view, where steepening occurs at the high-density regions found around the dense cores.

  By plotting the mass-weighted density exponent
 \begin{equation}
  k_{\rho, \rm mean} = \int \frac{k_{\rho} \rho {\rm d} v}{\rho {\rm d} v} \;,
 \end{equation}

  against the mean density (Fig. \ref{fig:overview}), we find that regions of higher densities tend to have steeper density profiles, which suggests a \emph{density-driven collapse} scenario.
  Since all regions belong to the Perseus molecular cloud, we assume that they have almost the same age $t_{\rm Perseus}$. Provided that they evolve at paces set by their free-fall time $t_{\rm ff}$, which is related to the mean densities by  $t_{\rm ff}\approx
  \sqrt{1 / (G \rho_{\rm mean, init})}$, the evolutionary stage can be parameterized using $f = t_{\rm Perseus}/t_{\rm ff} \sim \rho_{\rm mean, init}^{1/2}$. 
  Regions with larger $\rho_{\rm mean, init}$ are more evolved, have higher densities when observed (higher $\rho_{\rm mean, present}$),  with steeper density profiles and higher dense gas fraction.   In reality, due to the presence of turbulence, the actual collapse time can be a few times the collapse \citep{2018MNRAS.477.4951L,2022ApJ...927...75A}. But this additional factor will not change our conclusion.
  The picture of a steepening density profile is consistent with results from a recent paper \citep{2020arXiv200914151G}, and the correlations point to a simple picture where the mean gas density sets the pace at which a region evolves, leading to the steepening of the density profile \footnote{One might be worried about the effect of protostellar outflows \citep{2016ARA&A..54..491B}. Indeed, regions like the NGC1333  \citep{2000A&A...361..671K} contain a number of these. However, the effects should not be significant, as recent simulations found that outflows lead to an excess of low-density gas which we do not analyze \citep{2022ApJ...927...75A}. }.

  % Our result indicate that the evolution of molecular cloud is best characterized by a continuos steepening of the density profile. This trend has been 
  % found in simulations of \citet{2020arXiv200914151G}. To describe gravitational collapse,  some authors adopt the langue of ``phase transition'', and believed that a cloud should be divided into different regions \citep{}. In

% e evolution is reflected by their steeper density profiles (smaller $k_{\rho}$) and higher dense fractions. Cloud evolution can thus be described as the steepening of the density profile driven by gravity, and the rate is determined by the mean density of the gas. 

The map also allows us to identify a new region called   ``Perseus Cirrous"
(Fig. \ref{fig:overview}) for the first time. Although overlooked by previous
studies, the region stands out in our analyses due to its shallow density
profile ($k_{\rho } \approx -1$). The region occupies the shallow end of the
density exponent parameter space and its structure should be representative of
the structure gas at the early stages of gravitational collapse. On the other hand,
since this region has a density profile that is significantly shallower than
those the others, it is possible that this region belongs to the gravitationally
unbound phase, whereas others belong to the bound phase, as suggested by
previous studies \citep{2007ApJ...665..416K,2012ApJ...750...13C,Li2017}.

\subsection{Summary \& Future Extensions} 
% By incorporating levelset method into the analysis of the cloud structure, we are able to quantify the structure of a cloud in terms of the density exponent $k_{\rho}$ and provide a spatially resolved map of $k_{\rho}$. Our analysis have revealed the diversity of cloud structure, from which we propose that cloud evolution should be described by a process where the density profile steepens. 
% The density exponent map provides a new way to confront simulations with observations. Observationally, the density exponent map can be used to predict star formation in different environments. 
% By
% introducing the level-set method, we propose a framework to study the density structure of
% molecular clouds in terms of the density exponent $k_{\rho}$, which measures the steepness of the density profile measured in the log-log space. Through our analysis have revealed the diversity of density structure measured in terms of $k_{\rho}$. The fact that the density exponent is directly related to the free-fall time points to a unified picture where gravitational collapse leads to a continuous steepening of the density profile. Apart of this, the existence of regions with very shallow and steep density profiles deserves further explorations. 
The evolution of molecular clouds is exemplary of complex, multi-scale processes. Regions in molecular clouds appear to be gravitationally bound at 
$\sim $ pc scale \citep{2015A&A...578A..97L}, and the collapse of dense cores, which is directly related to star formation, occurs at  $\lesssim $ 0.05 pc.  Importing the Level-Set Method, we develop a new, robust formalism to compute spatially resolved maps of the density exponent.  On the pc scale,  the mass-weighted mean density exponent correlates with the star formation activity. On smaller scales  ($\lesssim$ 1 pc), the density exponent $k_{\rho}$ still exhibits significant variations. This complex pattern results from a continued steepening of the density profile driven by gravitational collapse.

 The spatial information our method provides is valuable for large,  inhomogeneous datasets.  The Level Set-based formalism can be modified to suit different models, for example, to derive the scale length for exponential-like structures,  which we will explore in the future.
 We expect our method to be effective for other structures including the density structure of granular materials and
 the Large-Scale Structure of the
 Universe.

 One limitation of our method is that it still requires 3D density. In the current approach, we are using 3D density distribution constructed from 2D surface density maps. For non-self-similar structures, the relation between exponent evaluated in 2D and 3D are not trivial \footnote{This has been well-understood under the Abel Transform \url{https://en.wikipedia.org/wiki/Abel_transform}.}, and our ongoing studies (Li \& Zhou in prep) suggest that compared to the current approach, evaluating the density exponent from 2D is less accurate but significantly easier to implement, making widespread applications feasible. These will be explored in our future papers.

%  The spatial information our method provides is valuable since the density structures of these clouds can be inhomogeneous.
\section*{Acknowledgements}
We thank the referee for a careful reading of the paper and for the comments. 
GXL acknowledges supports from 
NSFC grant W820301904 and 12033005. 

\section*{Data availability}
The data this publication makes use of is publicly available, the link can be found in  \citet{2016A&A...587A.106Z}, and we use code is from \citet{2022arXiv220105484L}, which is available at \url{https://gxli.github.io/Constrained-Diffusion-Decomposition/}.

\bibliographystyle{mnras}

\bibliography{paper}{}

\clearpage 

\appendix

\section*{Reconstruction of 3D density structure}\label{sec:appen:3d}
We develop a formalism to construct 3D density structures from 2D observations.ð
The reconstruction consists of two steps. First, using a method called
``constrained diffusion decomposition'' \citep{2022arXiv220105484L} , we
decompose the surface density maps into component maps that contain structures
of different sizes. The first of these maps contain structures whose sizes range
from 1 to 2 pixels, and the $n$th of these maps contain structures of sizes\
between $2^{n-1}$ and $2^n$ pixels.  Second, a 3D density structure is
constructed using the component maps. During our reconstruction, the channels
are assumed to be slabs of different thicknesses, where and along the line of
sight,  the $n$th channel has a Gaussian density profile of dispersion of
$2^n$ pixels. The final 3D density structure is assumed to be the sum of these
slabs. When combining these slabs, we aligned them such that the density
maximums stay on the same plane. 

Due to the lack of information on the distribution of gas along the line of
sight direction, the density structure we constructed is not identical to
but resembles the real one. We first test our method by producing a 3D clump of
where $\rho \sim r^{-2}$, projected it to 2D and verified that our
reconstruction allows us to recover the density exponent to an accuracy of
$\lesssim 0.01$. Then, using results from numerical simulations
\citep{2019MNRAS.486.4622C}, we perform density exponent analysis on both the
original data and the 3D data constructed from a 2D projection. They simulated the collapse of a 1000 $M_{\odot}$ clump, with a sophisticated treatment of the chemistry and thermal physics of the
ISM. The initial condition resembles regions that resemble the clumps found in Perseus. The simulation produced a density structure that is sufficiently complex, which enables us to test the robustness of our method.

\begin{figure}
  \includegraphics[width= 0.45 \textwidth]{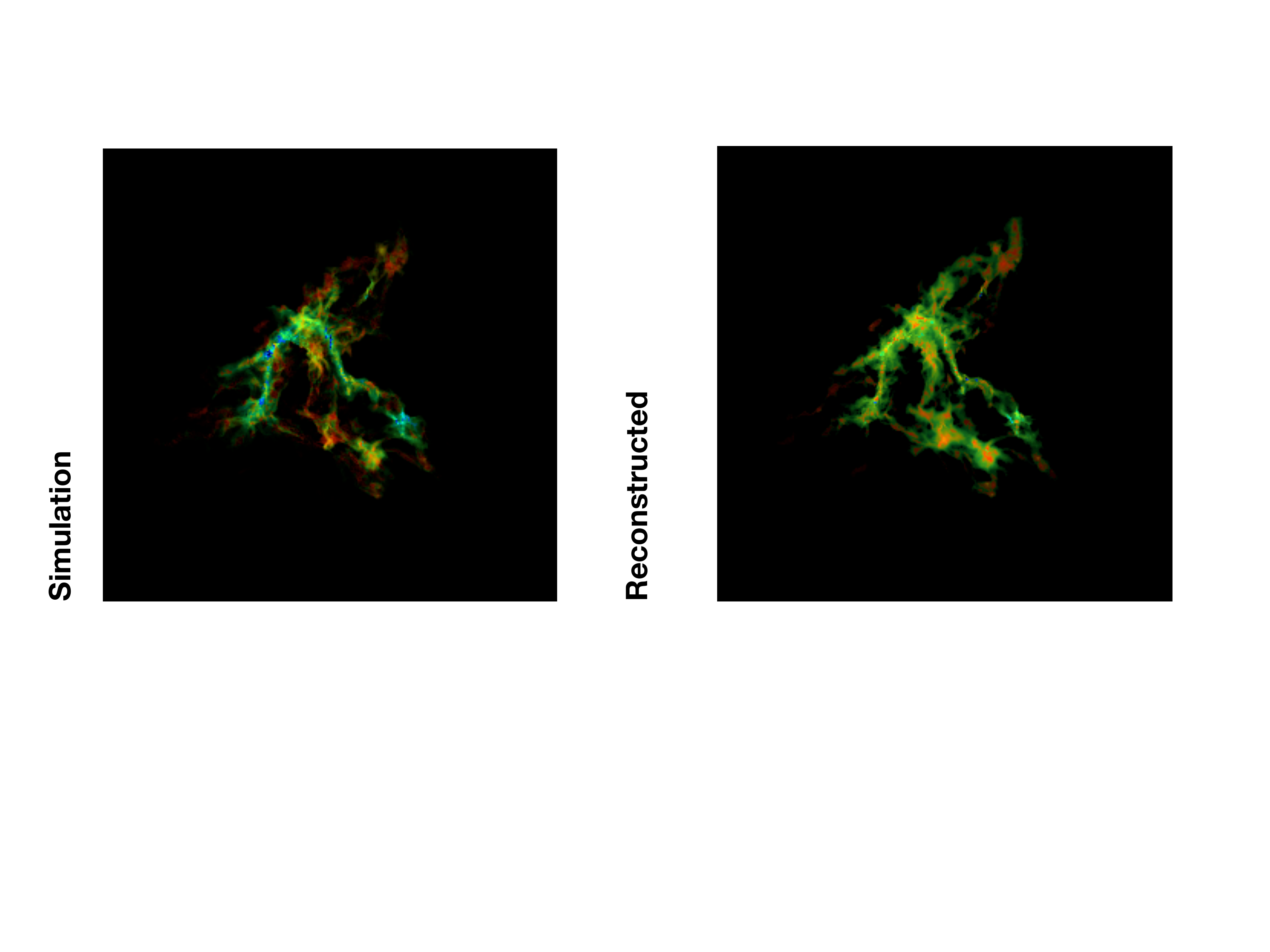}

\caption{\label{fig:trend} {\bf  Testing the 3D density reconstruction method.} {\bf Left panel:} Map of mass-weighted mean density exponent of a simulated cloud. {\bf Right panel:}  Map of  mass-weighted mean density exponent of the reconstructed cloud.  The brightness represents the density, and the color represents the density exponent. Red:  $k_{\rho} = -1 \pm 0.7 $, green:  $k_{\rho} = -2 \pm 0.7 $, and blue:  $k_{\rho} = -3 \pm 0.7 $. \label{fig:compare}}
\end{figure}

We limit ourselves to gas with $n({\rm H_2}) \gtrsim 250 \;\rm cm^{-3}$ and find that
the reconstructed cloud and the original cloud are similar in terms of
$k_{\rho}$ (Fig. \ref{fig:compare}). The original cloud has a mass-weighted
density exponent of -1.66, and the reconstructed cloud has $k_{\rho} = -1.69$. The difference is
noticeable but is still small compared to the variations we are interested in. Although some small-scale details are lost, compared to the original cloud, the reconstructed cloud has very similar density exponent
distributions. 

\section*{Measurement of dense gas fraction and star formation activity}
\label{sec:appendix:dense}
The star
formation activity is characterized by the dense gas ratio -- the ratio between
the amount of the  dense gas which should collapse to form stars and the total amount of gas in a
region. To trace the dense gas, we use the 870$\mu$m observation towards the  Perseus region where only dense gas can be observed \citep{2006ApJ...638..293E}, and created a mask containing significant ($\gtrsim 3.5\,  \sigma$) detections, which correspond to a surface density of  $ n_{\rm H_2} = 2 \times 10^{21}\;\rm cm^{-2} $ (Converted from a limiting sensativity of $3.5\times 15\rm  mJy\; beam^{-1}$, using the values \citealt{2008A&A...487..993K}, and this observations a 5-$\sigma$ mass-limit of 0.18 $M_\odot$). Using the Herschel-Planck surface density map \citep{2016A&A...587A.106Z}, the total amount of gas is derived by integrating  over whole regions, and the amount of dense gas
 is derived by integrating over subregions with significant  Bolocam 870$\mu$m detections.

 \section*{Relation between density exponent, surface density exponent and density power-law}\label{sec:c}
 Following previous papers \citep{2011ApJ...727L..20K,Girichidis2014,Li2016}, assuming a sphere whose density is 
\begin{equation}
  \rho \propto r^{k_\rho}\;, 
\end{equation}
where $k_rho < 0$, the (volume-weighted) density distribution is 
\begin{equation}
P_{\rm v, \rho} \propto r^2 {\rm d } r \propto \rho^{ 3/k_{\rho - 1} } {\rm d} \rho\ \propto \rho^{3 / k_\rho} {\rm d} \;{\rm log}(\rho)\;.
\end{equation}
Thus 
\begin{equation}
  {\rm log}( \frac{ P_{\rm v, \rho}}{ {\rm d }\;  ({\rm log} \rho)} ) \propto 3 / k_\rho   \;,
\end{equation}
where the slope density PDF is $3/k_\rho $. 

The corresponding surface density distribution is 
\begin{equation}
  \Sigma \propto r^{k_\rho} + 1\;, 
\end{equation}
thus 
\begin{equation}
  P_{\Sigma} \propto r {\rm d } r  \propto \Sigma^{2 / (k_\rho + 1) } {\rm d} {\rm log}(\Sigma) \;, 
  \end{equation}
  and 
\begin{equation}
  {\rm log}(\frac{P_{\rm \Sigma}}{{\rm d}  {\rm log} (\Sigma)}) \propto       2/(k_\rho + 1)\;,
\end{equation}
where the slope is $ 2/(k_\rho + 1)$.

\section*{Relation between density exponent and dense gas fraction}\label{sec:appendix:d}
\begin{figure}
  \includegraphics[width = 0.45   \textwidth]{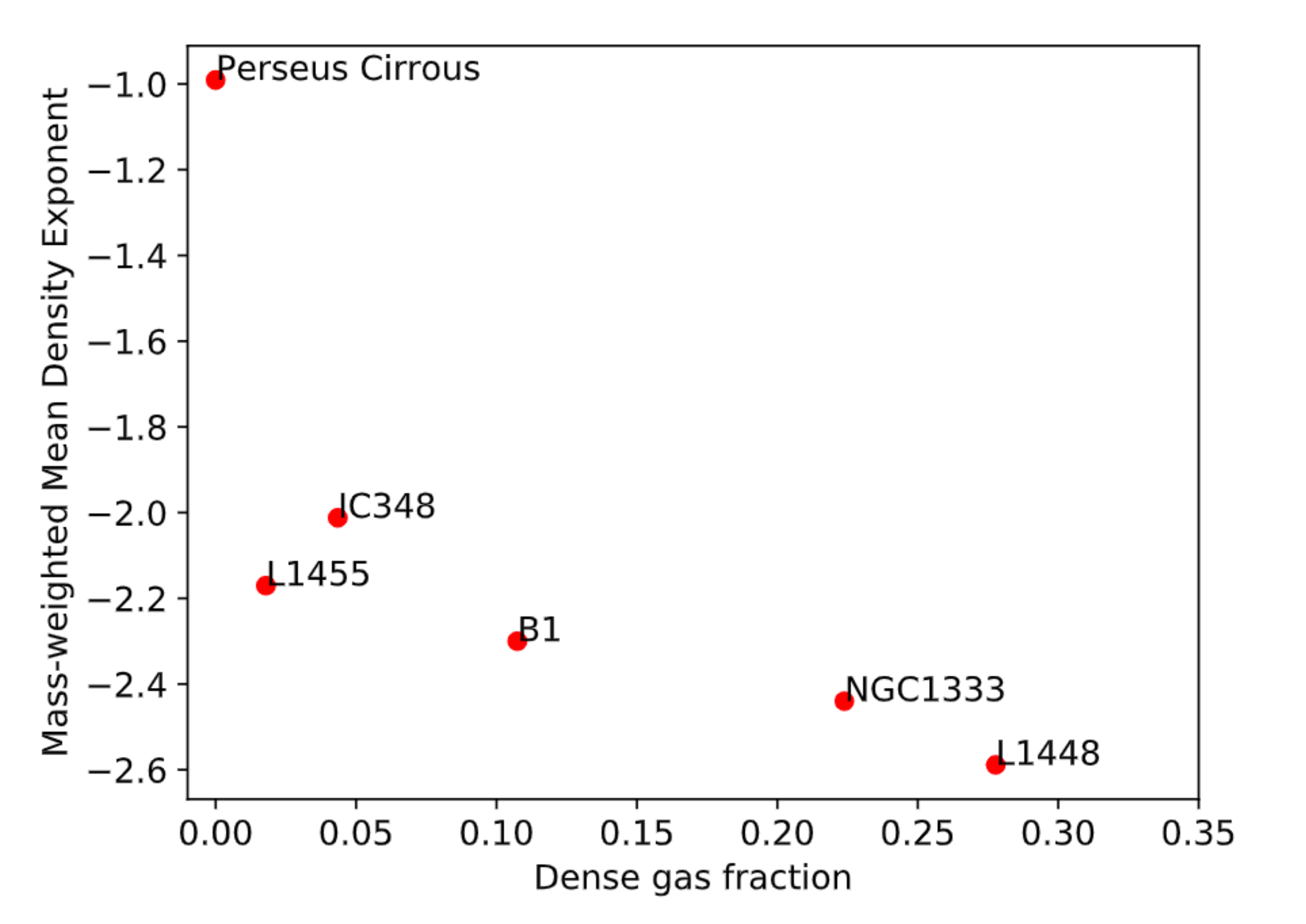}
  \caption{Relation between density exponent and dense gas fraction. \label{fig:pie} }
\end{figure}

In Fig. \ref{fig:pie} we plot the relation between density exponent and dense gas fraction.  

\end{document}